\documentclass[twocolumn,showpacs,preprintnumbers,amsmath,amssymb,prl]{revtex4}
\usepackage{graphicx}% Include figure files
\usepackage{dcolumn}% Align table columns on decimal point
\usepackage{bm}% bold math

\begin{document}

\preprint{APS/123-QED}l
\title{Fabry-Perot interference, Kondo effect and Coulomb blockade in carbon nanotubes}
%\title{Transition from closed to open carbon nanotube quantum dot}
\author{K. Grove-Rasmussen}
\email{k\textunderscore grove@fys.ku.dk}
\author{H. I. J\o rgensen}
\author{P. E. Lindelof}
\affiliation{Nano-Science Center, Niels Bohr Institute, University
of Copenhagen, Denmark}
\date{\today}% It is always \today, today,

\begin{abstract}
High quality single wall carbon nanotube quantum dots have been made
showing both metallic and semiconducting behavior. Some of the
devices are identified as small band gap semiconducting nanotubes
with relatively high broad conductance oscillations for hole
transport through the valence band and low conductance Coulomb
blockade oscillations for electron transport through the conduction
band. The transition between these regimes illustrates that
transport evolves from being wave-like transmission known as
Fabry-Perot interference to single particle-like tunneling of
electrons or holes. In the intermediate regime four Coulomb blockade
peaks appear in each Fabry-Perot resonance, which is interpreted as
entering the SU(4) Kondo regime. A bias shift of opposite polarity
for the Kondo resonances for one electron and one hole in a shell is
in some cases observed.
\end{abstract}

\maketitle
\section{Introduction}
Single wall carbon nanotubes (SWCNT) are interesting objects for the
study of low dimensional mesoscopic systems, where the observed
phenomena crucially depends on the coupling to the contacts. For
good contact, the SWCNT acts as an electron wave guide creating
resonances at certain energies. Such systems are regarded as open
quantum dots with the resonances corresponding to the broad energy
levels of the quantum dot \cite{Liang}. In the opposite limit of
very low transparency, the electrons are forced to tunnel one by one
due to Coulomb blockade and the energy levels sharpens due to their
longer life time \cite{Tans1998,BokrathQDropes1997}. In this
so-called closed quantum dot regime the electron number on the SWCNT
is well-defined except at charge-degeneracy points where single
electron tunneling occurs. An intermediate regime also exists in
which the electron number on the dot is still fixed but significant
cotunneling is allowed. This leads to different kinds of Kondo
effects related to the total excess spin
\cite{GoldhaberGordon1998Nature,NygaardKondo,Paaske} and/or the
orbital degree of freedom on the SWCNT quantum dot
\cite{JarilloOrbitalKondo2005Nature,Makarovski}. We will in this
paper examine the transition between these regimes in small band gap
semiconducting nanotubes, where the coupling to the SWCNT is
different for transport through the valence and conduction band
\cite{Cao2002NatureMaterials}. High quality measurement is presented
from each transparency regime showing that each Fabry-Perot
oscillation develops into four Coulomb blockade resonances with
finite conductance in the valleys for intermediate transparency
(Kondo effect) and well known Coulomb diamond patterns with
four-fold degenerate shell structure at lowest transparency
\cite{Sapmazprb}.

\vspace*{2.0cm}

\section{Experimental methods}
SWCNTs are grown from predefined catalyst islands by chemical vapor
deposition on a highly doped silicon substrate capped by a 500\,nm
silicon dioxide layer. After growth, pairs of electrodes are defined
by electron beam lithography some microns away from the catalyst
islands in hope that one SWCNT bridges the gap between the two
electrodes. In the case of the small band gap semiconducting SWCNT
the electrodes consist of Au/Pd bilayers (40\,nm/10\,nm). Finally
bonding pads of Au/Cr are made by optical lithography connecting the
electron beam lithography defined electrodes. The devices are
electrically probed at room temperature with typical resistances in
the range of 20-200\,k$\Omega$. These are cooled to low temperatures
where device characteristics clearly reveals if only one SWCNT is in
the gap. More details on device fabrication can be found in Refs.
\cite{MSS2006Proc,hij}. The measurements presented in this article
are mostly made at 4\,K in a DC-setup.

\section{Metallic and Semiconducting nanotubes}
\begin{figure}
\center
\includegraphics[width=0.48\textwidth]{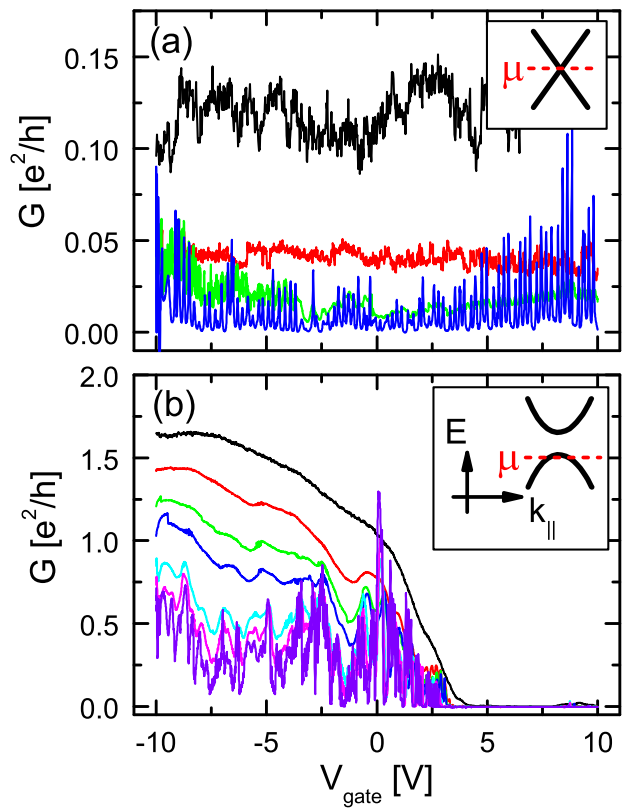}
\caption{Two different types of SWCNTs. (a) A metallic SWCNT
identified by its weak gate voltage dependence at high temperature
($T= 220, 120, 38, 10$\,K, black to blue). At low temperatures
oscillations in the conductance versus gate voltage are seen due to
Coulomb Blockade. Inset: Band diagram of a metallic SWCNT
(armchair). (b) A semiconducting SWCNT identified by its strong gate
dependence of the linear conductance. Oscillations of the
conductance at low temperature in the p-type region is due to single
hole transport (Coulomb blockade). Inset: Band diagram of a
semiconducting SWCNT with the electrochemical potential in the
valence band corresponding to the situation for $V_{gate} < 4$\,V.
Curves are taken at temperatures ($T=150, 70, 40, 30, 15, 10,
3.3$\,K, black to violet).} \label{fig1}
\end{figure}

SWCNTs have the remarkable property that depending on the exact
arrangement of the carbon atoms they can either be metals or
semiconductors. Figure \ref{fig1} shows the gate (and temperature)
dependence of the linear conductance for two different types of
SWCNTs, which defines whether the nanotube is semiconducting or
metallic. Figure \ref{fig1}(a) displays the behavior of a metallic
SWCNT characterized by its relatively constant linear conductance as
a function of gate voltage close to room temperature. On the
contrary the behavior shown in Fig.\ \ref{fig1}(b) corresponds to a
semiconducting SWCNT due to its strongly gate dependent linear
conductance \cite{TansTransistor}. The semiconducting SWCNT has
clearly high conductance for negative gate voltages, which is due to
hole transport through the valence band as indicated in the inset
showing the band structure (p-type). In the range of 4\,V
$<V_{gate}<10$\,V the chemical potential of the leads is aligned
with the band gap, i.e., no conduction occurs. At higher
temperatures signs of electron transport through the conduction band
is seen at high positive gate voltages due to thermal excitation of
electrons into the conduction band. The SWCNT shown in Fig.\
\ref{fig1}(b) has a relative large band gap, while conduction for
small band gap semiconductors reappears for positive gate voltage in
the gate range shown due to electron transport through the
conduction band (see below). Note, that the resistance between the
two devices differ by an order of magnitude due to different
coupling to the leads. In general the coupling can to some extent be
controlled by choice of electrode material \cite{Babic}.

Figure\ \ref{fig1}(a-b) also shows the temperature dependence, where
the gate dependence of the linear conductance for both devices
evolves into regular oscillations at low temperature. These
oscillations are a manifestation of Coulomb blockade, which as
mentioned above happens for SWCNT weakly coupled to the leads. The
regularity of the Coulomb oscillations indicates whether good
quality of the SWCNT has been obtained or if more than one SWCNT is
bridging the gap between the electrodes. For high quality SWCNTs
regular oscillations should persist through a gate region of
typically $|V_{gate}|<10$\,V as shown in (a). On the contrary if the
SWCNT has defects only small regions of gate voltage might have well
behaved oscillations.

\section{Small band gap semiconducting nanotubes}
We will now focus on small band gap semiconducting SWCNTs, where the
band gap is so small that transport can be tuned from hole transport
through the valence band to electron transport through the
conduction band by the back gate
\cite{Jarillo-Herreroe-hsym,CaoSmall,Cao2004PhRvL,Cao2002NatureMaterials}.
\begin{figure*}
\center
\includegraphics[width=0.7\textwidth]{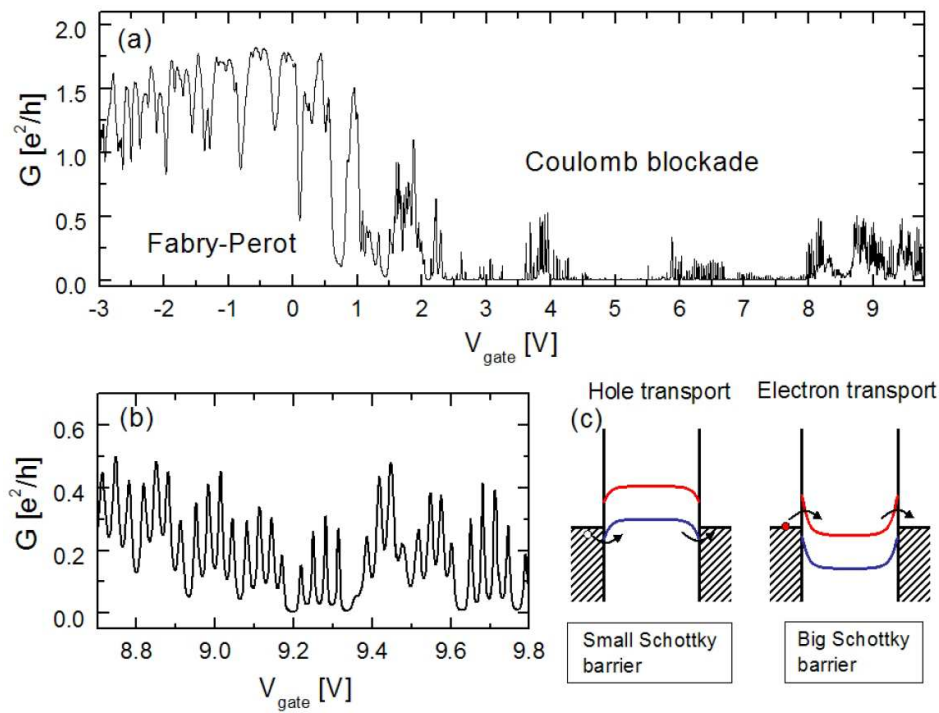}
\caption{Measurements at 4\,K of a small band gap semiconducting
SWCNT (device 1). (a) Linear conductance versus gate voltage. For
hole transport through the valence band at negative gate voltage
high conductance Fabry-Perot oscillations are observed, while
electron transport through the conduction band at positive gate
voltage is dominated by Coulomb blockade due to higher Schottky
barriers. (b) Zoom at high positive gate voltages where the
conductance increases and Coulomb blockade resonances are spaced in
four. (c) Schematic band diagrams of a small band gap single wall
carbon nanotube contacted to leads, where the band bending is
controlled by the gate voltage. The red/blue band is the
conduction/valence band, respectively. Left: The condition for hole
transport through the valence band. Holes tunnel into/out of the
valence band through a relative small Schottky barrier. Right:
Condition for electron transport through the conduction band, where
electrons tunnel into/out of the conduction band through a larger
Schottky barrier. Thus high conductance is observed through the
valence band in contrast to low conductance through the conduction
band, {\em i.e.}, Fabry-Perot versus Coulomb blockade regime. }
\label{fig2}
\end{figure*}
Figure \ref{fig2}(a) shows the linear conductance versus gate
voltage at source-drain voltages $V_{sd} \sim 1$ m$e$V for a small
band gap SWCNT at $T=4$\,K (device 1). The nature of the SWCNT is
identified by the relatively high conductance region for negative
gate voltages (hole transport) in contrast to the low conductance
region for positive gate voltages (electron transport). The broad
oscillations for hole transport for gate voltages between $-3$\,V
and 0.5\,V are a manifestation of the Fabry-Perot interference
pattern, {\em i.e.}, an open quantum dot \cite{Liang}. In contrast
for positive gate voltages regular low conductance oscillations are
observed due to Coulomb blockade. Figure \ref{fig2}(b) shows the
high positive gate region for electron transport where relatively
high conductance Coulomb blockade resonances are seen. They are
spaced into four reflecting the spin and orbital degree of freedom.
When this device is cooled to lower temperature (50\,mK) Kondo
resonances are seen within the four-fold shell structure (not shown)
in contrast to the lower conducting region, e.g., gate voltages from
5\,V to 8\,V, where only single electron tunneling is possible.

The different transparency of the n- and p-type regions can be
understood from Fig.\ \ref{fig2}(c). For negative gate-voltages
(left figure) the bands are bending in such a way that holes can
tunnel from source into the valence band and out to drain. The
Schottky barrier for hole transport is relatively small because the
workfunction of the Pd contacts is close to the valence band edge
leading to a relatively high conductance
\cite{Cao2002NatureMaterials}. Transport can be changed to electron
transport through the conduction band (right figure) by applying
positive voltage to the gate. The Schottky barrier is in this case
significantly larger leading to a low coupling of the SWCNT to the
electrodes. Between these two transport regions no transport is
allowed because the chemical potential in the leads is within the
band gap of the semiconducting SWCNT.
\begin{figure*}
\center
\includegraphics[width=0.8\textwidth]{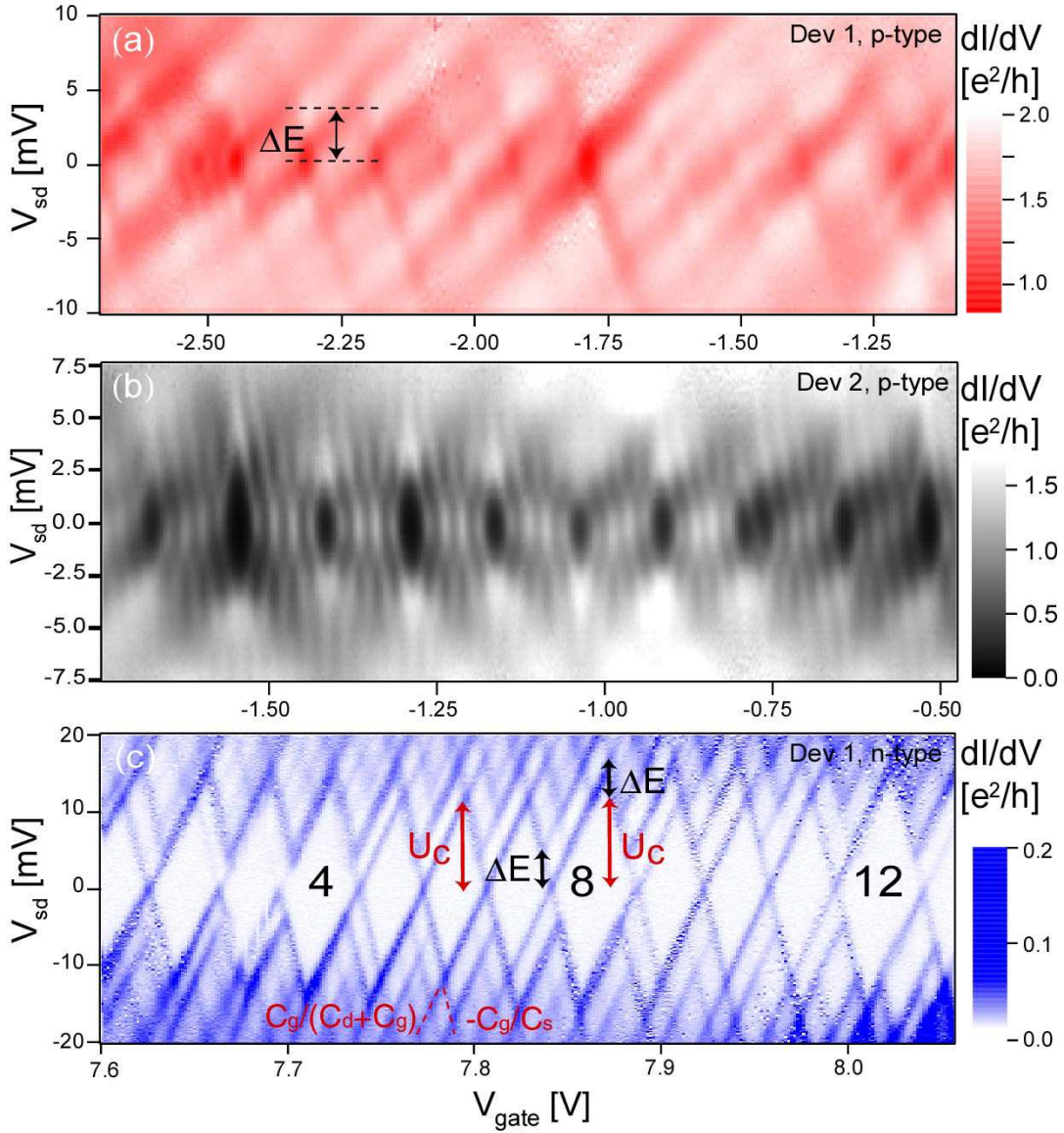}
\caption{Measurements at 4\,K of two different small band gap SWCNTs
with Au/Pd contacts (device 1 and 2). (a) Bias spectroscopy plot of
a small gate region of device 1 in the p-type gate region of Fig.\
\ref{fig2}(a) showing a Fabry-Perot interference pattern, i.e., an
open quantum dot. The level spacing is estimated to $\Delta E\sim
4$\,m$e$V. (b) Single hole transport is visible on top of the broad
Fabry-Perot resonances when the transparency is lowered (device 2,
p-type) by four Coulomb blockade peaks splitting each resonance. The
four peaks with finite conductance in the valleys is due to SU(4)
Kondo effect. (c) For even lower transparency of device 1 clear
Coulomb blockade diamonds are seen. The four-fold degeneracy is
clearly observed and the charging energy and level spacing can be
extracted as shown by the red and black arrows, respectively. The
dashed red lines indicate the slope used to estimate the
capacitances of the device in n-type region.} \label{fig3}
\end{figure*}

More information on the transport properties can be revealed by bias
spectroscopy, where the differential conductance is measured as a
function of gate and bias voltages. A bias spectroscopy plot of part
of the Fabry-Perot region is shown in Fig.\ \ref{fig3}(a). The low
conductance regions (red areas) form a mesh due to interference of
the hole waves reflected back and forth. Regarding the device as a
quantum dot the white regions at zero bias correspond to being in
resonance, while a red regions are off resonance. The level spacing
can be extracted as indicated in Fig.\ \ref{fig3}(b) by the black
arrow $\Delta E \sim 4 $\,m$e$V consistent with the device being
around $L= \frac{\hbar v_F \pi}{\Delta E} \sim 400$\,nm. Here we
have used a linear dispersion since we are far from the band gap.
Furthermore, a four-fold degeneracy of the level is assumed, which
is clearly revealed in the n-type region of the device (see below).
The device has a high asymmetric resistive coupling, because the
conductance at resonance is lower than $4e^2/h$. The asymmetry
$\Gamma_L/\Gamma_R$ can be found from
$G_{res}=\frac{4e^2}{h}\frac{4\Gamma_L
\Gamma_R}{(\Gamma_L+\Gamma_R)^2}$, where $G_{exp,res}\sim2e^2/h$ is
the conductance at resonance yielding $\Gamma_L/\Gamma_R=0.17$. The
capacitive couplings to the source and the drain electrodes are also
slightly different or/and the capacitive coupling to the gate is
comparable to the source and drain capacitance because of the
different slopes of the low conductance lines (red). The
capacitances will be examined more closely in the Coulomb blockade
case below.

Figure \ref{fig3}(b) shows a bias spectroscopy plot in the p-type
region for another small band gap semiconducting SWCNT with
lower transparency (device 2). Signs of Fabry-Perot oscillations are still
observed, but each Fabry-Perot resonance is split into four smaller
peaks \cite{Cao2002NatureMaterials}. These peaks are due to Coulomb
blockade and single hole tunneling, {\em i.e.}, the four-fold
degenerate level becomes visible due to quantization of the charge. The finite conductance in the valleys between the peaks are
indication of SU(4) Kondo effect, which will be treated in more detail below \cite{JarilloOrbitalKondo2005Nature}.

Figure \ref{fig3}(c) shows a bias spectroscopy plot in the n-type
region of device 1, where the transparency is reduced even further.
Clear Coulomb blockade diamonds and excited states are observed. The
numbers indicate the relative electron filling of the SWCNT quantum
dot and a number dividable by four corresponds to a filled shell
identified as three small diamonds followed by a bigger diamond. A
shell thus consists of a four-fold degenerate level as expected due
to orbital and spin degrees of freedom. The charging energy can be
extracted from the three small diamonds as half the source-drain
height $U_c \sim 11$\,m$e$V (left red arrow). The orbital splitting
and exchange energy are very small since the three consecutive small
diamonds in one shell are almost equal in height, e.g., diamonds 5,
6 and 7. Every fourth diamond is bigger (filled shell) since the
addition of the first electron in a new shell requires both a
charging energy and a level spacing energy. The additional diamond
height of the larger diamonds thus yields the level spacing $\Delta
E \sim 4 $\,m$e$V (right black arrow). This is consistent with the
level spacing identified from the excited state lines shown for
electron filling 7 (left black arrow). The asymmetry of the diamonds
are due to the capacitive coupling of the source $C_s$, drain $C_d$
and gate $C_{g}$ electrodes to the SWCNT. Estimating the slopes
$\alpha_s=0.51$ and $\alpha_d=-1.02$ of the two lines constituting
the diamond (dashed red lines) as well as the gate voltage distance
$\Delta V_g=33.6$\,mV between two Coulomb diamonds (diamond 7) the
capacitances can be found. We find $C_s=4.7$\,aF, $C_d=4.6$\,aF and
$C_{g}=4.8$\,aF by $\alpha_s=C_g/(C_g+C_d)$, $\alpha_d=-C_g/C_s$ and
$\Delta V_g = e/C_g$, where $C=C_s+C_d+C_{g}$ is the total
capacitance to the surroundings, $\alpha_{s/d}$ corresponds to
aligning the electrochemical potentials of the dot with the
source/drain and an asymmetric biasing with the drain on ground is
used \cite{Hanson}. The equal magnitude of the gate and source/drain
capacitive coupling thus makes the diamond asymmetric.
\begin{figure}[t]
\center
\includegraphics[width=0.48\textwidth]{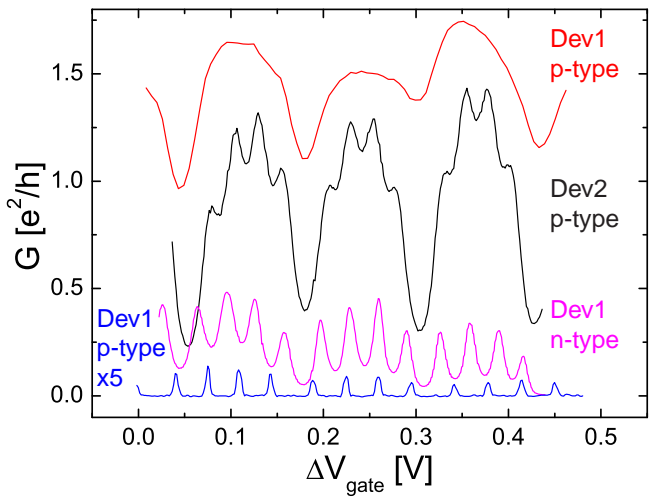}
\caption{Linear conductance at 4\,K,
where the behavior evolves from Fabry-Perot to Coulomb blockade
resonances because of the decreasing coupling to the leads. The red
curve shows broad Fabry-Perot resonances from Fig.\ \ref{fig3}(a)
with no single hole charging effects, while the four-fold
periodicity becomes visible due to Coulomb blockade for the black
curve from Fig.\ \ref{fig3}(b). This structure is even more evident
when the coupling decreases further as shown for the n-type region
of device 1 (magenta and blue curves). All the curves are translated along the
gate axis, but not scaled.} \label{fig:FPtoCB}
\end{figure}

The transition between the different transparency regimes is even
more clearly revealed in the linear conductance versus gate voltage
shown in Fig.\ \ref{fig:FPtoCB}. All curves are extracted at zero
bias from bias spectroscopy plots in the p- or n-type region of the
two small band gap semiconducting SWCNTs presented above (device 1
and 2). The most conducting device (red curve) shows broad
Fabry-Perot oscillations with no sign of single hole transport
consistent with holes added continuously to the SWCNT due to the
relatively good coupling. When the coupling to the SWCNT weakens,
the holes become more localized on the SWCNT and single hole
transport is observed (black curve). Four Coulomb blockade
resonances emerge in each broad Fabry-Perot resonance consistent
with each resonance (level) being four-fold degenerate. For even
weaker coupling a closed quantum dot is formed (magenta and blue
curves), where the measurement stems from the n-type region of
device 1. The curves are not scaled but only shifted along the gate
axis. Similar observations have been made by Cao {\em et al.} on
very clean \textit{suspended} small band gap semiconducting SWCNTs
and also in our measurement the Coulomb energy seem to diminish as
the transparency increases \cite{Cao2002NatureMaterials}. This is
seen by the distance between the peaks within one shell for the
magenta curve is smaller than in the case of the black curve.

\subsection{Intermediate regime}
\begin{figure}[t]
\center
\includegraphics[width=0.48\textwidth]{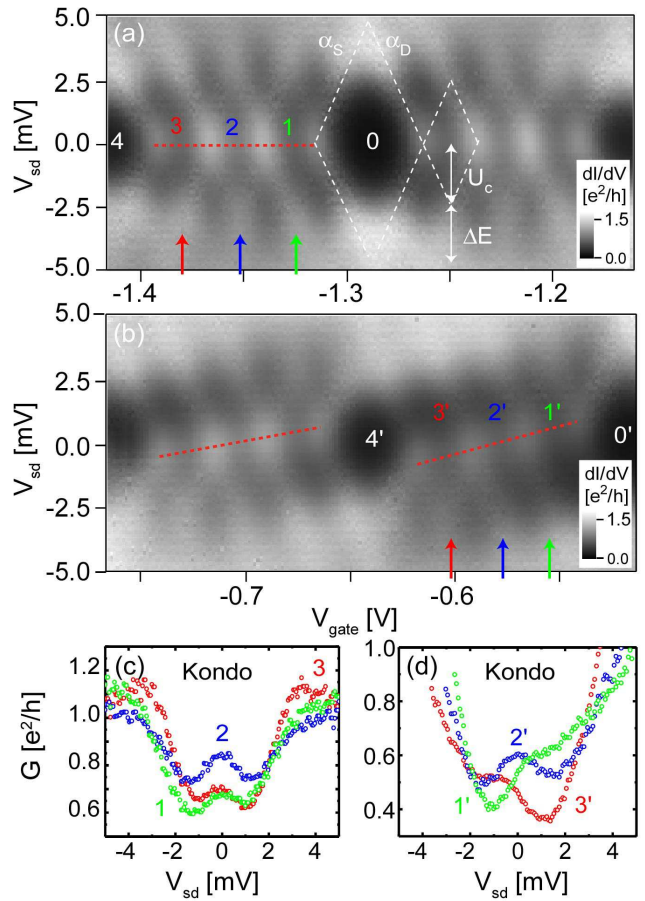}
\caption{(a-b) Bias spectroscopy plots at 4\,K of smaller gate regions taken from Fig.\ \ref{fig3}(b) showing SU(4) Kondo effect. Numbers indicate the number of holes in a shell.
The Coulomb diamonds are shown by white
 dashed lines and the white double arrows reveal the level spacing and the charging
 energy. The red dashed lines show how the Coulomb and Kondo resonances are shifted in bias as holes are added to the shell, i.e., no and finite
 slope for (a) and (b), respectively.
  The green, blue and red arrows point to bias cut for hole filling one, two and three displayed in (c-d). Clear Kondo peaks are observed,
   while they are shifted to opposite bias for hole filling 1' and 3' in (d).} \label{fig5}
\end{figure}

We now want to examine the transport behavior in the regime of
intermediate transparency in more detail. Figure \ref{fig5}(a) shows
a bias spectroscopy plot taken from Fig.\ \ref{fig3}(b) with the
filling of only two shells each containing a four-fold degenerate
level, i.e., addition of 8 holes as the gate voltage becomes more
negative. The numbers 0,..,4 show the number of holes in one of the
shells. The Coulomb diamonds are still faintly visible despite the
relative high transparency, where the big diamond(s) correspond to
filled shells. The charging energy and level spacing can be
estimated as above yielding $U_c\sim\Delta E\sim 2.5$\,meV (see
arrows). Similarly, to the case of the closed quantum dot we deduce
the capacitances of the device from the slopes of the diamond sides
($\alpha_d=-0.185 and \alpha_s=0.178$) and the voltage distance
between two consecutive Coulomb blockade resonances ($\Delta
V_g=27$\,mV) in the shell giving $C_s=32$\,aF, $C_d=27$\,aF and
$C_g=6$\,aF. In the case of the closed quantum dot of device 1, the
gate capacitance is almost identical ($\sim 5$\,aF) consistent with
the two devices having the same geometry. In contrast to the more
closed dot (device 1), the source and drain capacitances are
increased. We speculate that increasing the transparency to the
source and drain contacts leads to an increase of the source and
drain capacitances because the wave function extends below the
contact. Such behavior is also consistent with the charging energy
diminishing with increasing transparency as noted above. Figure
\ref{fig5}(c) shows the bias cuts in the center of the diamonds for
filling 1, 2 and 3 (see colored arrows), where clear zero bias peaks
are present. These zero bias peak are manifestation of the SU(4)
Kondo effect due to the four-fold degeneracy of the shell
\cite{JarilloOrbitalKondo2005Nature,ChoiPRL}. The FWHM of the Kondo
peak for filling of 2 holes yields 1.5\,mV by a Lorentzian fit
corresponding to a Kondo temperature of 5.1\,K higher than the
measurement temperature of 4K. The center positions of the Kondo and
Coulomb blockade resonances within a shell are at zero bias as
expected indicated by the red dashed line.

Figure\ \ref{fig5}(b) shows another gate region from Fig.\
\ref{fig3}(b) with SU(4) Kondo effect showing a different behavior
of the Coulomb blockade and Kondo resonance positions. The Coulomb
blockade resonances are clearly shifted to more negative
source-drain voltage as the holes are added to the shell with a
slope of $\Delta V_{sd}/\Delta V_{gate}\sim 0.008$ (red dashed
line). Furthermore, the Coulomb blockade resonances for the hole
transitions 0'-1' (1'-2') and 3'-4' (2'-3') are shifted equally in
bias but with different polarity. Similarly, the Kondo peak for
holes fillings 1' and 3' (i.e., one electron) are shifted oppositely
in bias ($\sim \pm 0.5$\,meV)  as shown in Fig.\ \ref{fig5}(d),
while
 the Kondo peak for hole filling 2' is centered at zero bias. We note that an opposite slope is also found in Fig.\ \ref{fig3}(b) ($V_{gate}\sim 1.1$\,V).
This behavior has previously been observed in carbon nanotubes
showing four-fold shell structure \cite{Makarovski,BabicKondo} and
also for spin half Kondo effect in a GaAs based quantum dot
\cite{Simmel}. In Ref.\ \cite{Makarovski} the authors speculate that
it is related to SU(4) Kondo effect, because SU(4) theory predicts a
shifted density of states compared to the spin half Kondo effect for
electron filling 1 in equilibrium \cite{ChoiPRB}. However, in Ref.\
\cite{BabicKondo,Simmel} and theoretical work on spin half Kondo
effect \cite{Kraviec} the behavior was attributed to the asymmetry
of the coupling to the leads, which shifts the Kondo peaks to finite
biases. The asymmetry can be extract from the linear conductance of
the Kondo resonance in the unitary limit or from current plateau
corresponding to the total current through one level
\cite{BabicKondo}. However, at this temperature the Kondo resonances
are not saturated and the current plateaus are too smeared to
extract the couplings. More experimental work at lower temperature
as well as precise theoretical predictions of these shifts are
needed to elaborate more on this effect.

\section{Conclusion}
In conclusion measurements on very clean single wall carbon nanotube
quantum dots have been presented. We focused on small band gap
semiconducting SWCNTs and the transition from an open to a closed
quantum dot (Fabry-Perot interference to Coulomb blockade). The
appearance of four peaks is observed in each Fabry-Perot resonance
as the transparency is decreased, which is interpreted as entering
the SU(4) Kondo regime. The Kondo resonances for one hole and one
electron in a shell are in some cases shifted to opposite biases. At
even lower transparency clear Coulomb blockade shell structure with
a four-fold degeneracy due to orbital and spin degrees of freedom is
observed.

\section{Acknowlegdement}
We would like to thank Jens Paaske for discussions and the support of the EU-STREP Ultra-1D
program.

\bibliography{PhysicaE}% Produces the bibliography via BibTeX.
\end{document}